**Operando probing of nanocracking in CuO-derived Cu during CO₂ electroreduction**

Jiawei Wan[1,2,11], Ershuai Liu[3,4,11], Woong Choi[5,6,11], Jiayun Liang[2], Buyu Zhang[1,2], Keon-Han Kim[3,4], Xianhu Sun[1], Meng Zhang[7], Han Xue[1], Yi Chen[1,2], Qiubo Zhang[1], Changlian Wen[3], Ji Yang[5], Karen C. Bustillo[8], Peter Ercius[8], Denis Leshchev[9], Ji Su[5], Zakaria Y. Al Balushi[1,2], Adam Z. Weber[4,5], Mark Asta[1,2], Alexis T. Bell[3,4,10], Walter S. Drisdell[3,4*], Haimei Zheng[1,2*]

[1] Materials Sciences Division, Lawrence Berkeley National Laboratory, Berkeley, CA 94720, USA

[2] Department of Materials Science and Engineering, University of California, Berkeley, Berkeley, CA 94720, USA

[3] Chemical Sciences Division, Lawrence Berkeley National Laboratory, Berkeley, CA 94720, USA

[4] Liquid Sunlight Alliance, Lawrence Berkeley National Laboratory, Berkeley, CA 94720, USA

[5] Energy Storage and Distributed Resources Division, Lawrence Berkeley National Laboratory, Berkeley, CA 94720, USA

[6] Department of Energy Engineering, Gyeongsang National University, Jinju, Gyeongnam, 52828, Republic of Korea

[7] California Institute for Quantitative Biosciences, University of California, Berkeley, Berkeley, CA 94720

[8] National Center for Electron Microscopy, The Molecular Foundry, Lawrence Berkeley National Laboratory, Berkeley, CA 94720, USA

[9] National Synchrotron Light Source II, Brookhaven National Laboratory, Upton, NY 11973, USA

[10] Department of Chemical and Biomolecular Engineering, University of California, Berkeley, Berkeley, CA 94720, USA

[11] These authors contributed equally: Jiawei Wan, Ershuai Liu, Woong Choi

[*] Corresponding authors: H.Z. Email: hmzheng@lbl.gov, W.D. Email: wsdrisdell@lbl.gov

**Abstract**

Identifying and controlling active sites in electrocatalysis remains a grand challenge due to restructuring of catalysts in the complex chemical environments during operation[1–3]. Inactive precatalysts can transform into active catalysts under reaction conditions, such as oxide-derived Cu (OD-Cu) for $CO_2$ electroreduction displaying improved production of multicarbon ($C_{2+}$) chemicals[4–6]. Revealing the mechanism of active site origin in OD-Cu catalysts requires in situ/operando characterizations of structure, morphology, and valence state evolution with high spatial and temporal resolution[7,8]. Applying newly developed electrochemical liquid cell transmission electron microscopy combined with X-ray absorption spectroscopy, our multimodal operando techniques unveil the formation pathways of OD-Cu active sites from CuO bicrystal nanowire precatalysts. Rapid reduction of CuO directly to Cu within 60 seconds generates a nanocrack network throughout the nanowire, via formation of "boundary nanocracks" along the twin boundary and "transverse nanocracks" propagating from the surface to the center of the nanowire. The nanocrack network further reconstructs, leading to a highly porous structure rich in Cu nanograins, with a boosted specific surface area and density of active sites for $C_{2+}$ products. These findings suggest a means to optimize active OD-Cu nanostructures through nanocracking by tailoring grain boundaries in CuO precatalysts. More generally, our advanced operando approach opens new opportunities for mechanistic insights to enable improved control of catalyst structure and performance.

## Main

Cu is the only metallic catalyst for the $CO_2$ electroreduction reaction ($CO_2$RR) that produces significant multicarbon ($C_{2+}$) chemicals[9–11]. Among various Cu-based materials, the oxide-derived Cu (OD-Cu) exhibits enhanced electrocatalytic activity and it is easy to prepare at low cost for potential practical applications[4–6]. Although diverse types of grain boundaries[12,13] and residual $Cu^+$ species[5,14] from precatalyst reconstruction have been reported as the active sites, the formation mechanisms and design principles for OD-Cu catalysts remain lacking. A major hurdle is the inability to directly monitor the structural evolution of OD-Cu catalysts under operating conditions[7] due to the fast, complex dynamic transformation behavior of precatalyst activation during $CO_2$RR[6,15]. Although advancements have been made recently in in situ/operando microscopy and spectroscopy studies of Cu-based catalysts during electrocatalytic reactions[16–22], tracking both morphological and compositional evolutions with high spatiotemporal resolution is needed to resolve the origin and evolution of OD-Cu active sites.

Here we report operando studies of OD-Cu catalysts evolution from CuO bicrystal nanowires during $CO_2$RR using a multimodal platform coupling the newly developed high-resolution electrochemical liquid cell transmission electron microscopy (EC-TEM) with time-resolved and high energy resolution fluorescence detected (HERFD-) X-ray absorption spectroscopy (XAS). We discover the formation pathways of catalytically active sites through nanocracking of CuO bicrystal nanowire precatalysts during rapid reduction to metallic Cu (Fig. 1a). Electrocatalytic performance testing in reactors of different scales (EC-TEM liquid cell, H-type cell, and membrane electrode assembly (MEA)), coupled with complementary ex situ structural characterizations, suggest the nanocracking behavior is general across all relevant operating conditions and is critical for enhanced $C_{2+}$ activity.

## CuO bicrystal nanowire precatalyst

CuO nanowires are prepared using a thermal oxidation method[23], with an optimized purification procedure. (Supplementary Fig. 1 and see Methods). TEM measurements show the typical morphology of a CuO nanowire with the growth direction of [1$\bar{1}$0] and bicrystal structure with the (002) twin plane in the middle (Fig. 1b and Supplementary Fig. 2 and 3)[24,25]. High-resolution TEM (HRTEM) image confirms its monoclinic crystalline structure with the (1$\bar{1}\bar{1}$) and ($\bar{1}\bar{1}$1) lattice ($d$ spacing ~2.51 Å) mirror reflected along the (002) twin plane ($d$ spacing ~2.52 Å) (Fig. 1c). X-ray absorption near-edge structure (XANES) of Cu K-edge of the as-prepared CuO nanowires and corresponding references demonstrate dominant CuO phase in the precatalysts (Fig. 1d). The Fourier transform of extended X-ray absorption fine structure (FT-EXAFS) reveals the characteristic peaks at 1.5 Å and 2.4 Å, which correspond to the Cu-O and Cu-Cu scattering paths in CuO, similar to the CuO standard spectra (Fig. 1e). Assisted by multifarious techniques (Supplementary Fig. 4-6), we conclude that CuO bicrystal nanowires are obtained as precatalysts for CO₂RR.

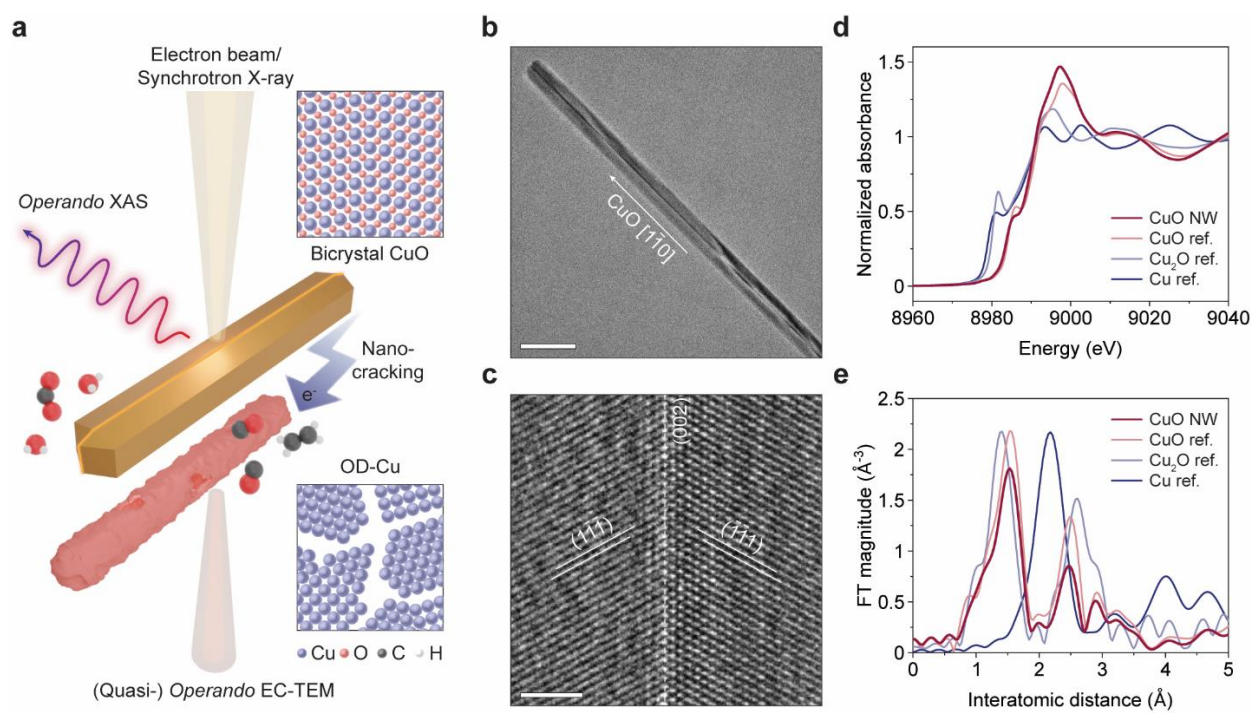

**Fig. 1 | The operando electrocatalytic platform used to study structural dynamics of nanocatalysts for CO₂RR and characterizations of CuO bicrystal nanowire precatalysts. a,** Schematic illustration of the operando electrocatalytic platform employed in this study to track the dynamic behaviors of CuO precatalysts under CO₂RR conditions. Quasi-operando EC-TEM uncovers the morphological and local structural changes of individual working catalysts (TEM setup limits direct detection of catalytic products), while operando XAS unveils the valence state evolution of the ensemble catalysts. **b,** TEM image shows the typical morphology of an as-prepared CuO bicrystal nanowire, which is divided by the twin boundary parallel to the nanowire's growth direction. Scale bar, 100 nm. **c,** HRTEM image of the local area on a CuO bicrystal nanowire, showing the atomic twinned structure. Scale bar, 2 nm. **d,** XANES and **e,** FT-EXAFS spectra of Cu K-edge for the as-prepared CuO nanowire (NW) precatalysts and the corresponding references (CuO, Cu₂O, and Cu), revealing characteristic structure of CuO phase.

**Visualizing nanocracking in catalyst restructuring**

We track the CuO nanowire precatalyst evolution under $CO_2RR$ conditions by quasi-operando EC-TEM measurements using our newly developed high-resolution electrochemical liquid cell[26]. Both top and bottom chips of the liquid cell are insulated and covered with electron transparent polymer membranes, while the interdigital Pt electrodes are deposited onto the bottom chip as the working and counter electrodes (WE and CE) (Fig. 2a, see Methods, and Supplementary Fig. 7-9). The as-synthesized CuO nanowires are drop-cast onto the Pt electrodes, followed by loading liquid electrolyte ($CO_2$-saturated 0.1 M $KHCO_3$) and liquid cell assembly. It can form multiple thin (<100 nm) liquid pockets encapsulated by the thin polymer membranes (Fig. 2a, Supplementary Fig. 10-12), which allows imaging of electrocatalyst evolution with unprecedented structural and compositional details with atomic/nanoscale resolution[26,27]. The cell potential is calibrated by the underpotential deposited hydrogen peak in 0.1 M perchloric acid in the electrochemical liquid cell[28], indicating -0.28 V vs. Pt is correlated to 0 V vs. standard hydrogen electrode (SHE) (Supplementary Fig. 13). A cell potential of -2.0 V vs. Pt (-1.32 V vs. reversible hydrogen electrode (RHE)) is applied on the CuO nanowires conditions in quasi-operando conditions that we simulate the $CO_2RR$ electrocatalytic reactions without directly measuring the reaction products (Supplementary Fig. 14-16).

The CuO bicrystal nanowire dramatically changes its structure and generates nanocracks while being reduced to Cu in 30 seconds under the operating conditions, which is observed from EC-TEM (Extended Data Fig. 1 and Supplementary Video 1). The formation and propagation of nanocracks within the CuO nanowire precatalysts under $CO_2RR$ conditions follow two different pathways: "boundary nanocracking" along the twin plane, and "transverse nanocracking" perpendicular to the surface/twin plane, as illustrated in Fig. 2b. Sequential images from the EC-TEM video show structural dynamics of an individual CuO nanowire on the Pt WE at the edge-on position with high spatiotemporal resolution (Extended Data Fig. 1, Fig. 2c, Supplementary Fig. 17, and Supplementary Video 2). Nanocracks are highlighted using a ridge detection method[29] and overlaid on the false-colored images of the nanowire, which clearly show the evolution and distribution of both boundary and transverse nanocracks.

Distinct characteristics of the nanocracking dynamics during CuO precatalyst electroreduction are observed, as shown in Fig. 2c. First, both boundary and transverse nanocracks are initiated immediately after applying potential, manifesting as nanocavities (0 s). Then, the transverse nanocracks propagate from the surface to the interior of the nanowire (8.5 and 10.0 s), while the boundary nanocracks emerge and expand as nanocavities continuously coalescing on the twin plane (26.0 s), as depicted in Fig. 2b. Finally, the transverse nanocracks encounter the larger/wider boundary nanocracks at the middle of the nanowire (30.0 s) as CuO is fully reduced, introducing an OD-Cu structure with a nanocrack network. Continuously, evolution of the nanocracks involves rich dynamic fluctuations of expansion and contraction (e.g. The boundary nanocrack dramatically opens at 50.0 s and partially heals at 60.0 s in Extended Data Fig. 1 and Supplementary Fig. 18), which occur constantly via an intricate process of OD-Cu catalysts self-reconstructions[16,17,19,22].

As the electroreduction reaction proceeds, a rapid increase in nanocrack density (estimated by the measured total area of nanocracks divided by the area of the nanowire at a given reaction time) is observed, starting from 0.04 to 0.08 within 10 seconds, and reaching a relatively stable stage of 0.10 in 30 seconds (Fig. 2d). It is worth noting that the propagation of the CuO/Cu phase boundary (from surface to the center) (Fig. 2c) and the shrinking volume (~30%) from CuO to Cu shown in the EC-TEM results (Supplementary Fig. 19) indicates the phase transition during the electroreduction of CuO. The formation of gas bubbles at catalyst surface is also observed via EC-TEM (Extended Data Fig. 1 and Fig. 2c), which can be attributed to the formation of gas phase species like carbon monoxide and hydrogen under the applied potential, indicating the on-going electrocatalytic reactions[9,10].

In situ energy dispersive X-ray spectroscopy (EDS) and HRTEM measurements are performed on the CuO nanowire-derived Cu catalysts inside liquid cell after the EC-TEM imaging without introducing artifact from sample transfer, providing complementary information on the real active species. The high-angle annular dark-field (HAADF) image and integral Cu and O elemental intensity profiles from corresponding EDS mapping reveal the boundary nanocrack within OD-Cu nanostructure, with no detectable signal of oxides (CuO and $Cu_2O$) (Fig. 2e and Supplementary Fig. 20). The HRTEM image and corresponding fast Fourier transform (FFT) pattern also confirm the dominant metallic Cu phase that is fully reduced from CuO precatalyst (See standard diffractions in Supplementary Fig. 6). The atomic structures show diverse Cu nanograins (single/poly-crystal) separated by the porous nanocrack network (Fig. 2f).

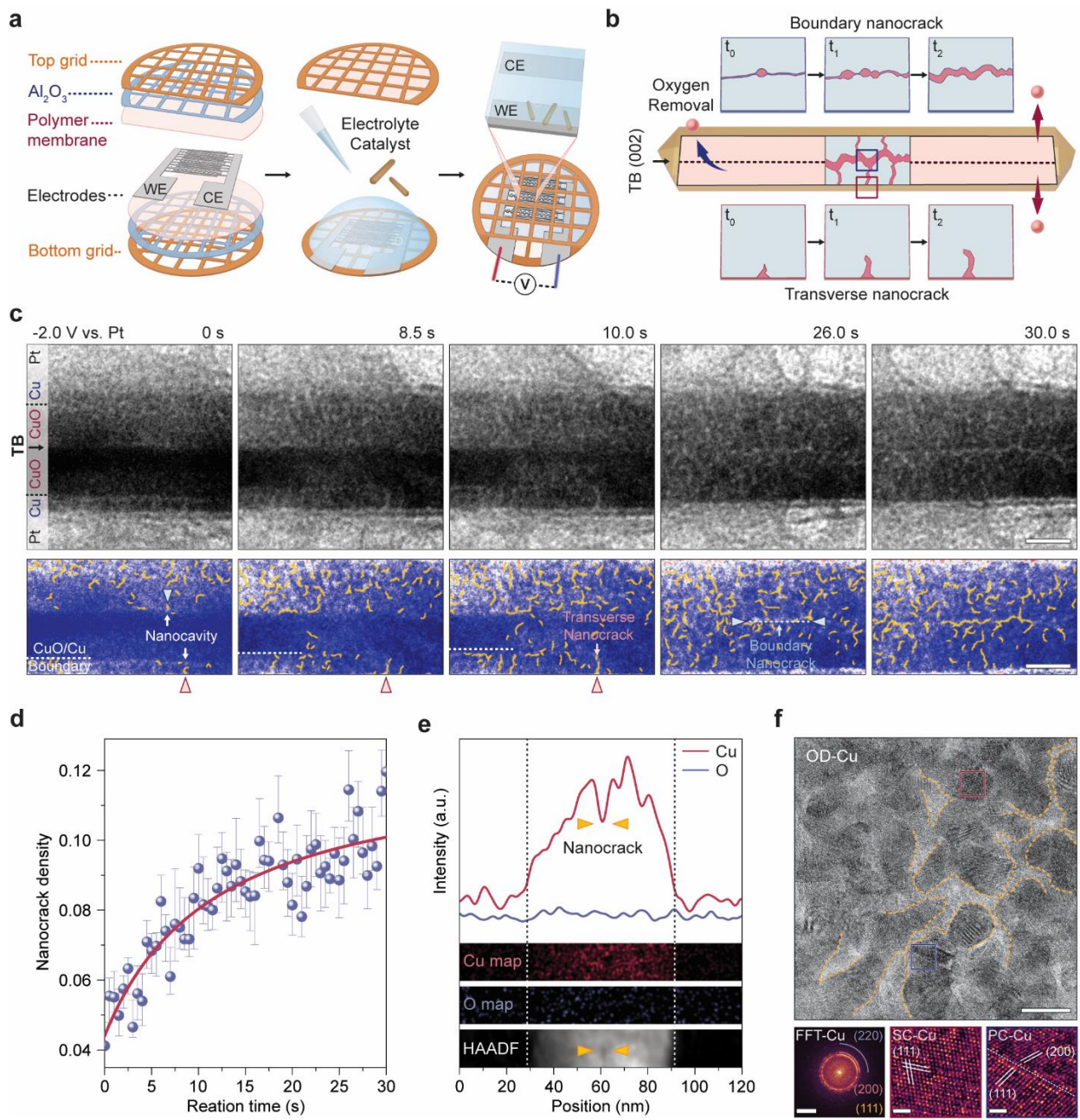

**Fig. 2 | Real-time probing of nanocracking in CuO bicrystal nanowire precatalysts under CO₂RR conditions using quasi-operando EC-TEM. a,** Schematic of fabrication and assembly processes of an EC-TEM liquid cell with the capability of enabling electrochemical measurements and simultaneously tracking the structural changes at nanoscale resolution. **b,** Schematic illustration of nanocracking within a CuO nanowire during electroreduction, boundary (blue box) and transverse (red box) nanocracks grow and propagate on a projected plane consisted with EC-TEM. **c,** Selected quasi-operando EC-TEM video frames of the morphological evolution of a CuO bicrystal nanowire at -2.0 V vs. Pt (-1.32 V vs. RHE) under CO₂RR

conditions, and corresponding distribution of nanocracks (highlighted in yellow) on false-color images of CuO nanowire (derived Cu). Transverse (red arrow) and boundary nanocracks (blue arrow), and CuO/Cu phase propagation (white dash line) are observed. Scale bar, 20 nm. **d,** Nanocrack density evolution in the first 30 s under $CO_2RR$ conditions. Systemic error may be introduced from detection under different contrast thresholds. **e,** In situ elemental analysis of CuO nanowire-derived Cu in c after restructuring. HAADF image, Cu, and O maps, and corresponding intensity profiles of Cu K-edge and O K-edge show existence of metallic Cu and nanocracks. **f,** In situ HRTEM image of cracked Cu after EC-TEM measurements. Scale bar, 10 nm. Nanocracks are highlighted in yellow dash line. FFT rings match with Cu (111), (200), and (220) facets, indicating dominated Cu phase. Scale bar, 5 $nm^{-1}$. Enlarged areas represent single crystalline (SC, red box) and polycrystalline (PC, blue box) Cu nanograins. Scale bar, 1 nm.

The restructuring mechanisms of CuO precatalyst during reduction (Fig. 2b) are further studied with density functional theory (DFT) and Nudged Elastic Band (NEB) method (Supplementary Fig. 21 and see Methods), showing preferential O removal on the boundary plane, leading to boundary nanocracks formation (Extended Data Fig. 2). Among the different diffusion pathways in CuO, it is energetically favorable for O to be removed perpendicular to the (002) twin boundary plane, which creates the transverse nanocracks. The reconstructed OD-Cu consists edges, corners, and grain boundaries (Fig. 2f), all of which may improve the $CO_2RR$ performance[12,13]. The connection between the wide center boundary nanocrack and the numerous smaller transverse nanocracks that compose the nanocrack network results in a highly porous structure with increased specific surface area and exposure of catalytic active sites, as well as providing efficient channels for reactant/product transport[30]. It may also improve activity through confinement effects[31,32].

**Tracking ensemble valence state evolution**

Since the nanocracking process is induced by reduction of the CuO bicrystal nanowire precatalysts into metallic Cu, we employ operando time-resolved XAS to comprehensively study the reduction kinetics and valence state evolution across a range of electrochemical conditions including $CO_2RR$ (see Methods). These measurements are highly sensitive to the ensemble average oxidation state, and are conducted in an H-type cell similar to those typically used for catalytic testing[33] (Supplementary Fig. 22). First, time-resolved XANES/EXAFS is applied for CuO nanowires at selected potentials based on the linear sweep voltammetry (LSV) (Supplementary Fig. 23). All the potentials are converted to the RHE scale with *iR* correction unless noted otherwise. At -0.65 V ($CO_2RR$ onset potential), initial reduction of CuO to Cu is observed within 30 s and is more dramatic after 60 s, indicated by the evolution of characteristic XANES features (pre-edge and white line intensity) (Fig. 3a and b) as well as diminishment of the Cu-O scattering path at 1.5 Å and emergence of Cu-Cu path of metallic Cu phase at 2.2 Å in FT-EXAFS (Fig. 3c). Around 120 s, the ensemble valence state stabilizes, closely matching the spectra at 1 h. It is noteworthy that the FT-EXAFS peak at 4.0 Å, representing the second shell of Cu-Cu scattering in metallic Cu, also increases in intensity with time, indicating that metallic Cu is becoming the dominant species after 120 s. At -1.0 V ($C_{2+}$ production potential), the complete reduction of CuO occurs within 60 s (Fig. 3d and 3e). The FT-EXAFS spectrum at 90 s under -1.0 V shows a negligible signal of Cu-O scattering peak at 1.5 Å, unlike the prominent shoulder peak observed at 90 s under -0.65 V, indicating a more thorough metallic phase formation (Fig. 3f). At all catalytically relevant potentials, the reduction of CuO occurs within 120 s, and the dominant species afterwards is metallic Cu (Supplementary Fig. 24-26, and Supplementary Tables 1 and 2). We also find more negatively below the thermodynamic CuO electroreduction potential can lead to a faster reduction (Fig. 3g). This fast reduction kinetics is consistent with quasi-operando EC-TEM measurements under -1.32 V on an individual nanowire, showing nanocracking and OD-Cu phase propagation, implying the CuO is fully reduced to Cu within 30 s (Fig. 2c). This suggests that the nanocracking behavior observed with EC-TEM also occurs at all catalytically relevant potentials in an H-type cell.

Next, we investigate whether the CuO nanowire ensemble reduces directly to metallic Cu, or whether the reduction pathway includes intermediate oxides (e.g. $Cu_2O$). For this we employ operando HERFD-XAS, which has increased energy resolution that increases sensitivity to small changes in XANES pre-edge features[34] (Supplementary Fig. 27 and 28). Sequential potentials between 0.45 V and -1.0 V are applied, with HERFD-XAS collected after current density stabilized at each potential (2-3 minutes) (Fig. 3h). Above 0.25 V, the CuO nanowire ensemble shows stable XANES features that match the CuO standard spectra (Supplementary Fig. 28), indicating that reduction is not yet triggered. At 0.15 V, the decrease in the major white line peak (8997 eV) and growth of the pre-edge peak (8980 eV) indicates the phase transformation to metallic Cu (Supplementary Fig. 29 and 30). The observed transition to metallic Cu is consistent with the formation of Cu nanograins observed with EC-TEM (Fig. 2f). Significant evolution of the edge feature at 0.05 V (Fig. 3h inset) indicates significant metallic Cu contributions. The 1st derivative XANES also shows an unambiguous major peak shift from 8984 eV (0.15 V) to 8979 eV (0.05 V) (Fig. 3i)[18], indicating the direct reduction of CuO into metallic Cu, which occur before the onset potential of CO₂RR catalysis. As the applied potential enters the CO₂RR regime (down to -1.0 V), more complete reduction is shown on the timescale of the HERFD-XAS measurement (Fig. 3h and i). It is noteworthy that small Cu⁺ signal is also found (8980.5 eV) in the 1ˢᵗ derivative XANES at open-circuit voltage (OCV) (Fig. 3i), which could be ascribed to the residual Cu₂O in the as-prepared CuO precatalysts[23] (Supplementary Fig. 6). However, this signal does not increase, suggesting that Cu₂O is not formed as an intermediate from CuO, and disappears below 0.05 V through full reduction towards metallic Cu.

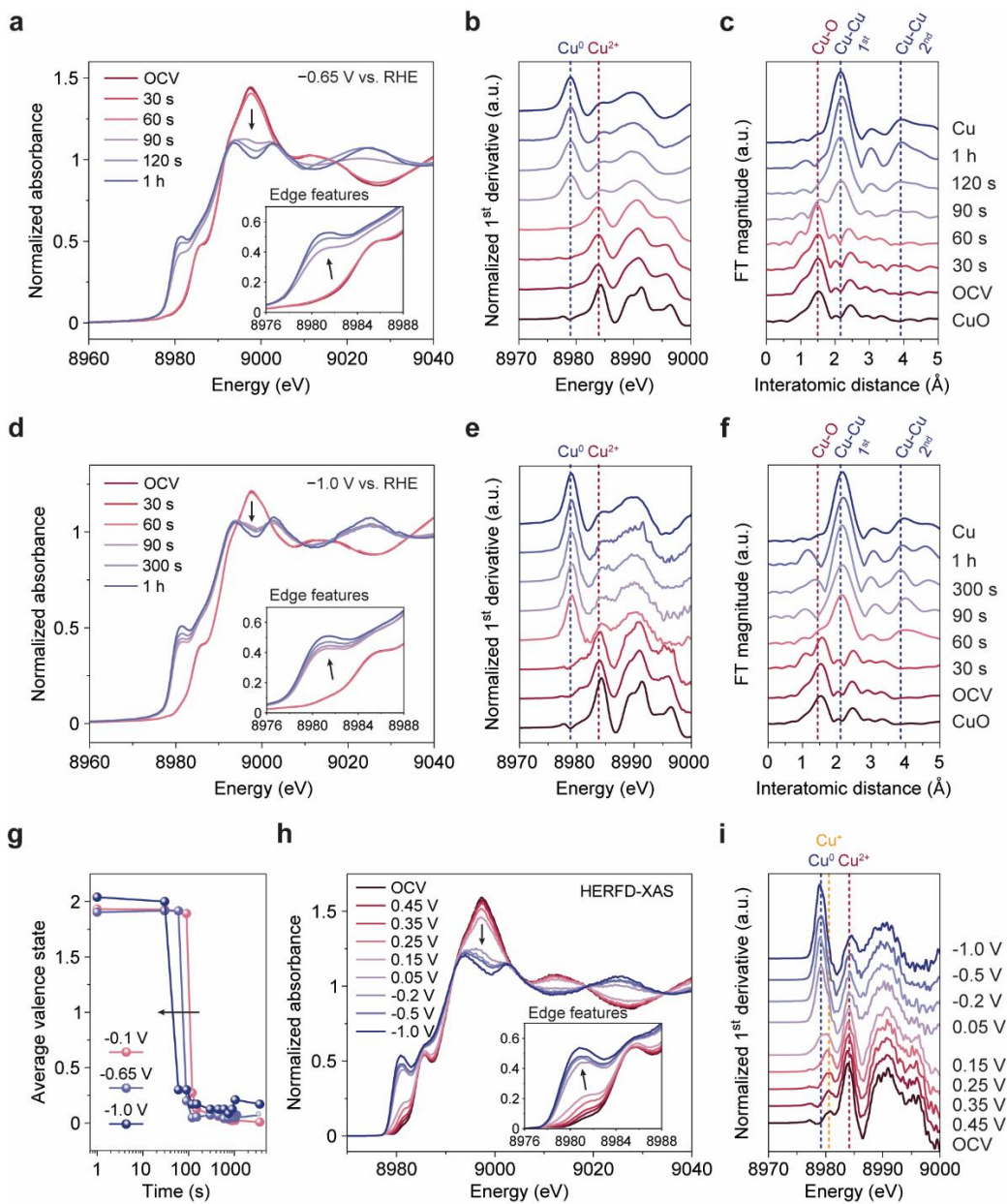

**Fig. 3 | Operando time-resolved (HERFD-) XAS studies of ensemble CuO bicrystal nanowire precatalysts under CO₂RR conditions. a,** Operando time-resolved conventional XANES spectra, **b,** corresponding 1$^{st}$ derivative spectra, and **c,** FT-EXAFS spectra of CuO nanowire precatalysts at -0.65 V vs. RHE under CO₂RR conditions. **d,** Operando time-resolved conventional XANES spectra, **e,** corresponding 1$^{st}$ derivative spectra, and **f,** FT-EXAFS spectra of CuO nanowire precatalysts at -1.0 V vs. RHE under CO₂RR conditions. **g,** Ensemble valence state evolution at different potentials. **h,** Potential dependence of operando Cu K-edge HERFD-XANES spectra of the CuO nanowire precatalysts under CO₂RR in 0.1 M CO₂-saturated KHCO₃ using chronoamperometry (CA) for 2-3 minutes and **i,** the corresponding 1$^{st}$ derivative spectra.

There is currently some debate in the field regarding the active sites for oxide-derived Cu catalysts for CO₂RR, with some studies claiming that such catalysts are fully reduced to metallic Cu during operation[13,35,36], and others positing that Cu⁺ species are stabilized in the catalyst and serve as active sites for C₂₊ product formation[5,14]. The specific behavior may depend on the precise state of the precatalysts in question. We propose that the reduction of CuO precatalysts is easier than that of Cu₂O, with a lower apparent activation energy, and it reduces directly to metallic Cu without formation of an intermediate or suboxide because the system can reach metastable states instead of forming Cu₂O[37,38]. The boundary of the bicrystal CuO nanowire provides additional oxygen diffusion pathways, which also boosts the reduction kinetics (Extended Data Fig. 2 and Supplementary Fig. 21). This is consistent with our conclusion that the active sites in the CuO nanowire-derived Cu catalyst are metallic Cu sites formed from the nanocracking process induced by the pristine oxide reduction (Fig. 2 and 3). We also note that signatures of oxide reappear quickly after the potential is released (Supplementary Fig. 31 and 32), emphasizing the need for operando studies to identify the valence state during operation.

**Structure-performance correlation in CO₂RR**

To establish the correlation between our observed catalysts structural dynamics and CO₂RR performance, we conduct electrolysis of the CuO nanowire precatalysts in CO₂-satuated 0.1 M KHCO₃ aqueous electrolyte in the same H-type cell design[33] used for XAS measurements. Different precatalyst loadings (0.05 to 1 mg cm⁻²) are evaluated under ambient conditions (see Methods and Supplementary Fig. 33). The stepwise chronoamperometry measurements reveal that the optimal potential for C₂₊ chemical production is -1.0 V for low loading samples of 0.05 and 0.1 mg cm⁻² (Supplementary Fig. 34 and Supplementary Table 3). The maximum Faradaic efficiency for the $C_{2+}$ production ($FE_{C_{2+}}$) and $C_{2+}/C_1$ product ratios are recorded as ~43% and 1.8 for a catalyst loading of 0.1 mg cm$^{-2}$ (Fig. 4a). However, both $FE_{C_{2+}}$ and $C_{2+}/C_1$ ratios decrease significantly for loadings above 0.2 mg cm$^{-2}$ (Fig. 4b and Supplementary Fig. 35). The decrease in $C_{2+}$ product selectivity at higher precatalyst loadings can be explained by the mitigation of CuO reduction due to thicker and less conductive layers, as demonstrated in the post-mortem TEM measurements (Supplementary Fig. 36). The precatalyst loading effect implies that the electroreduction of pristine oxide is indispensable for activating OD-Cu for $C_{2+}$ selectivity by creating electrocatalytic active sites[5,6,13,39]. To evaluate the initial process of catalyst activation and its effects on $CO_2RR$ performance, the electrochemically active surface area (ECSA) and roughness factor of the CuO nanowire-derived Cu catalysts are measured over time for 0.1 mg cm$^{-2}$ loading at -1.0 V (Fig. 4c, Supplementary Fig. 37, and Supplementary Table 4). Both the ECSA and roughness factor of the OD-Cu catalysts gradually increase to double their starting values in 30 minutes of operation, concomitant with an improvement of $CO_2RR$/HER FE ratio. This indicates that, following the initial nanocrack network formation within 30 seconds (Fig. 2c), the OD-Cu continues to restructure on a longer timescale, which is consistent with the frequently reported Cu restructuring during $CO_2RR$[16,17,19,22]. As a control experiment, we test pure Cu nanowire catalysts performance in the same H-type cell under 0.1 mg cm$^{-2}$ catalyst loading, and find the FE for $C_{2+}$ products (9.4%) significantly lower than that of the OD-Cu (43%) (Supplementary Fig. 38 and Supplementary Table 5).

The three-dimensional (3D) morphology of the CuO nanowire-derived Cu catalysts after H-type cell measurements, obtained by cryogenic transmission electron tomography (cryo-ET), shows a nanowire with highly roughened surface (Fig. 4e, Supplementary Fig. 39, Supplementary Video 3, and see Methods). Segmentation from the 3D structure in Fig. 4e displays the interior nanocrack network morphology (Fig. 4f). The nanocracks appear to be larger than EC-TEM results in Fig. 2, suggesting that propagation of nanocracks and nanocavities has occurred during Cu restructuring. Nevertheless, both boundary and transverse nanocracks are still maintained and observed in the final structure. Additional TEM and EDS confirm the existence of nanocracks (Supplementary Fig. 40-43). By contrast, pure Cu nanowires display less surface roughening after operation, and no nanocracks (Supplementary Fig. 44). This suggests that the initial nanocrack formation from reduction of CuO precatalysts provides an important framework for further OD-Cu restructuring that yields more active sites for $CO_2RR$ towards $C_{2+}$ products.

We note that Cu₂O is frequently found from our ex situ TEM experiments (Supplementary Fig. 43), but is not detected in the operando EC-TEM (Fig. 2f and g) or XAS (Fig. 3) at catalytically relevant potentials. This can be attributed to the inevitable re-oxidation of Cu catalysts as soon as they are out of operating condition. Indeed, we observe the formation of Cu$^+$ by operando XAS after releasing the cathodic potential (Supplementary Fig. 32), similar to our previous studies[21]. This underscores the importance of in situ/operando studies for accurate identification of the catalyst active sites.

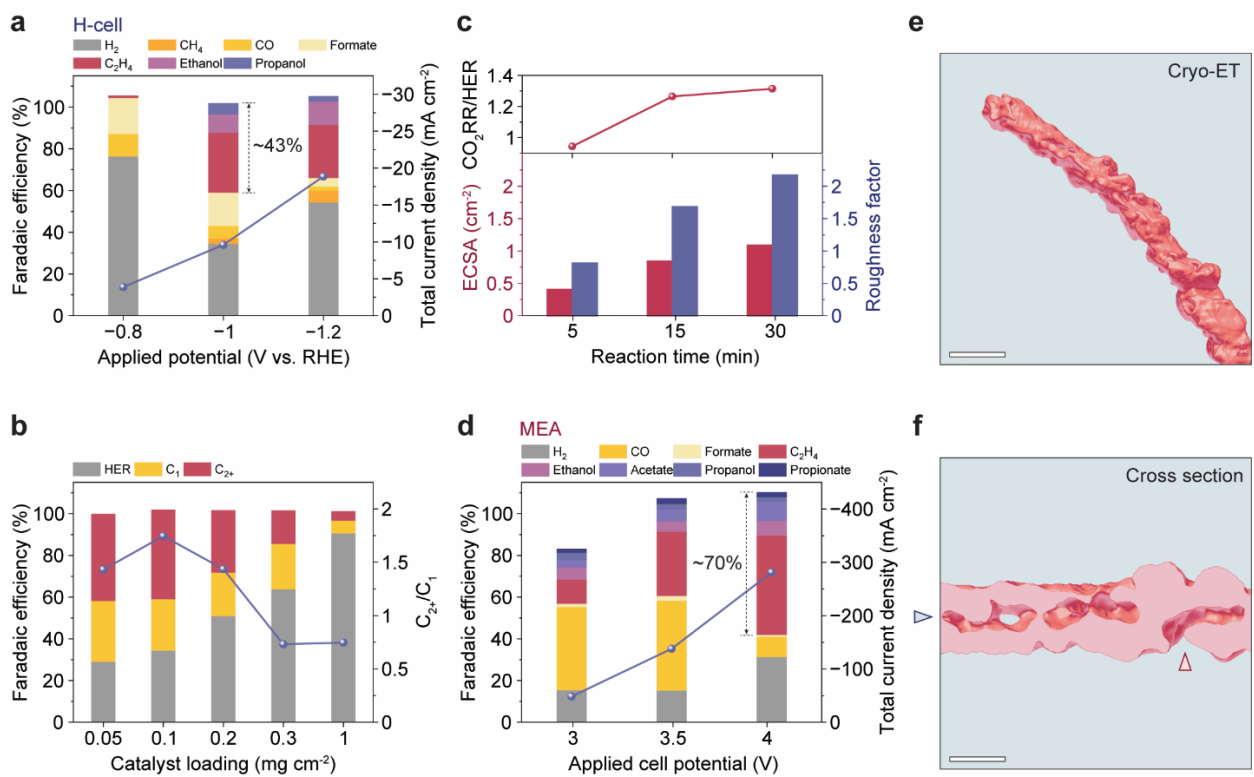

**Fig. 4 | Electrochemical CO₂RR performance and structural characterizations of CuO nanowire-derived Cu catalysts. a,** FEs for all CO₂RR products and total current density obtained from ensemble of CuO nanowire-derived Cu catalysts with precatalyst loading of 0.1 mg cm⁻² as a function of the applied electrode potential in an H-type cell. **b,** FEs for all CO₂RR products grouped as C₂₊ products, C₁ products and H₂ obtained from ensemble catalysts as a function of the precatalyst loading at -1.0 V vs. RHE in an H-type cell. **c,** Ratio of CO₂RR and HER, ECSA, and roughness factor of as a function of reaction of time with catalyst loading of 0.1 mg cm⁻² at -1.0 V vs. RHE in an H-type cell. **d,** FEs for all CO₂RR products and total current density obtained from ensemble catalysts as a function of the applied cell potential in an MEA electrolyzer. **e,** Cryo-ET reconstruction of an individual CuO nanowire-derived Cu nanoparticle obtained after CO₂RR performance testing in an H-type cell. Scale bar, 100 nm. **f,** Segmentation of an

enlarged area in e showing the porous interior structure with both boundary (blue arrow) and transverse (red arrow) nanocracks. Scale bar, 50 nm.

To further evaluate the application potential of CuO nanowire-derived Cu catalysts for industrial $CO_2RR$, we measure their catalytic performance in a 5 cm$^2$ MEA electrolyzer, which is the next generation electrolysis system (see Methods and Supplementary Fig. 45). The CuO nanowire-derived Cu exhibits ~70% $FE_{C2+}$ selectivity at the optimal current density of ~300 mA cm$^{-2}$ at 4 V (total cell voltage) in the MEA (Fig. 4d, Supplementary Fig. 46-48, and Supplementary Table 6). Such $C_{2+}$ selectivity improvement compared to H-type cells can be ascribed to the working environment that facilitates thorough reduction and activation of CuO inside MEA[40]. The superb performance of CuO nanowires in the MEA with low cost and easy preparation highlights their potential towards $CO_2RR$ in industrial applications. Post-mortem TEM measurements of OD-Cu (Supplementary Fig. 49 and 50) suggest the nanocracking behavior is general across all catalytically relevant conditions and reactors, including the EC-TEM liquid cell, H-type cell, and MEA.

In summary, this study systematically uncovers the dynamic structural reconstruction of CuO nanowire-derived Cu, from individual nanowires to ensemble catalysts, through advanced multimodal operando techniques under $CO_2RR$ working conditions. The OD-Cu structures are rich in nanocracks that dictate further dynamic restructuring, generating active sites for $C_{2+}$ products. The formation of nanocrack network during reduction of CuO bicrystal nanowire suggests a means to control catalyst structure. For example, polycrystalline CuO rich in diverse types of grain boundaries can be used as starting materials to generate an intricate architecture of predictable enriched boundary and transverse nanocracks during $CO_2RR$, boosting the performance by increasing the catalytic active sites. Similar strategies could also apply to other catalytic systems, such as sulfide/halide-derived Cu or others. This work demonstrates an innovative solution towards spatiotemporally resolving the complex and dynamic nature of precatalysts in electrochemical systems, and emphasizes the crucial relevance of applying complementary operando methods such as electron microscopy and X-ray spectroscopy for understanding structure-performance correlations.

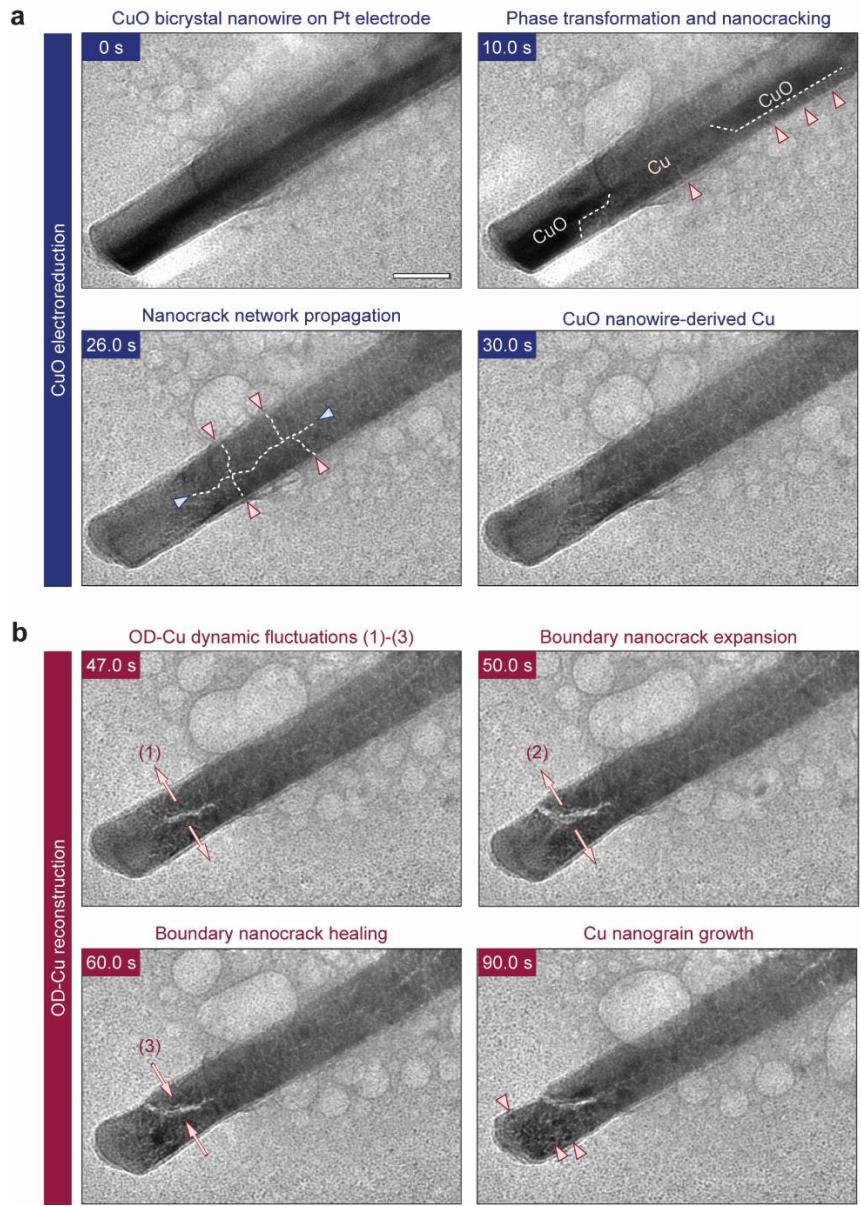

**Extended Data Fig. 1 | Quasi-operando EC-TEM monitoring electroreduction of an individual CuO bicrystal nanowire and the reconstruction of OD-Cu.** Selected quasi-operando EC-TEM video frames show (a) morphological evolution and phase transformation of a CuO bicrystal nanowire at -2.0 V vs. Pt (-1.32 V vs. RHE) under CO$_2$RR conditions. A nanocrack network is generated while CuO being reduced to OD-Cu within 30.0 s. (b) The OD-Cu continuously reconstructs and the nanocracks involve rich dynamic fluctuations. The boundary nanocracks dramatically expand and open at 50.0 s while they partially heal at 60.0 s (red arrows). The reconstructed Cu nanoparticles are also found (90.0 s). Scale bar, 50 nm.

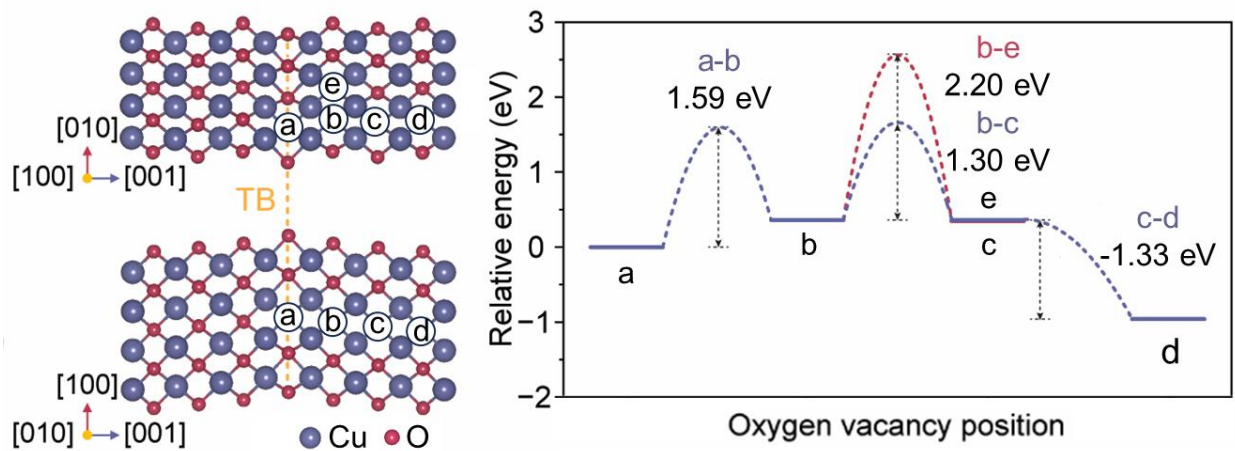

**Extended Data Fig. 2 | Theoretical calculations of O vacancy diffusion in CuO reduction.** DFT calculation results show energy of different O vacancy sites: (a) on the twin boundary (TB), (b) one atom away, (c) two atoms away perpendicular to twin boundary, and (d) on the surface. O vacancy prefer the sites at twin boundary and near surface, with relatively lower energy. This implies that O vacancies prefer to segregate at the boundary and will lead to the cracks along the twin boundary. NEB calculations further evaluate the energy barrier of different diffusion paths. We compare two paths: one is O vacancy diffuses perpendicular to the twin boundary (b)-(c) and the other one is O vacancy diffuses parallel to the twin boundary (b)-(e). The O vacancy diffusion barrier of (b)-(c) (1.3 eV) is smaller than that of (b)-(e) (2.2 eV), indicating O vacancy prefer to diffuse perpendicular to the twin boundary/surface, which can lead to formation of transverse cracks. Blue, red, and white balls represent Cu, O, and O vacancy, respectively.

**Methods**

**Preparation of CuO nanowire precatalysts**

CuO nanowires were synthesized through a typical thermal oxidation method[23] with an optimized purification procedure. The cleaned Cu substrates (0.1 mm Cu foils, ⩾99.9%, Thermo Scientific Chemicals) were heated at 450 °C for 90 minutes at ambient pressure, giving rise to the CuO nanowires growth on the substrates. CuO nanowires were further separated from the substrates by sonication in ethanol and purified by centrifuging. Additional synthesis details can be found in Supplementary Information.

**Fabrication of electrochemical liquid cells**

The electrochemical liquid cells for quasi-operando EC-TEM were built with a bottom and a top chip capsuling multiple liquid pocket in between. The bottom and top chips were fabricated based on the commercial Cu or Au TEM grids (200 square mesh, Electron Microscopy Sciences). The grids are first coated with 250 nm of electrically insulating aluminum oxide ($Al_2O_3$) by sputter deposition (AST-sputter, Berkeley Marvell NanoLab). Subsequently, a 10 nm-thick insulating polymer (0.25% Formvar solution in 1,2-dichloroethane (⩾99.0%, Sigma-Aldrich)) film was transferred onto the grids. Then, 10 nm-thick Pt interdigital electrodes were deposited onto the bottom chips using the e-beam evaporation (Semicore-evaporator, Molecular Foundry, LBNL) through a specially designed shadow mask, enabling electrochemical reaction in the micro-electrochemical system.

**Quasi-operando EC-TEM measurements**

Quasi-operando EC-TEM measurements were performed in $CO_2$-saturated 0.1 M $KHCO_3$ electrolyte within the self-fabricated electrochemical liquid cells using an FEI ThemIS aberration-corrected TEM at 300 keV (NCEM, LBNL). The EC-TEM videos were acquired at a speed of 500 ms per frame (2,048 × 2,048 pixels) at a beam dose rate of about 400 $e^-$ $Å^{-2}$ $s^{-1}$, which ensured the absence of beam-induced damage on the precatalyst materials. HAADF-STEM images and EDS were performed for chemical identification. Based on the multiple experimental results and analysis from different aspects, the EC-TEM measurements apply catalytically relevant potentials and environments. We are not able to directly detect the catalytic products during EC-TEM experiments, due to the limitation of equipment configuration, therefore, we consider the measurements as quasi-operando. Additional details and liquid cell assembly processes can be found in Supplementary Information.

**Operando XAS measurements**

The operando XAS experiments were conducted at room temperature in a homemade spectro-electrochemical H-type cell made by PEEK, with 0.1 M $KHCO_3$ electrolyte purged with Ar or $CO_2$. Before each measurement, the cell was held for 2-3 minutes to reach a pseudo-steady state. Data collection was then performed at the chosen potentials held during cathodic sweeps. Time-resolved experiments started at the same time when the applied potentials were turned on. All of the data were collected at Cu K-edge under fluorescence mode. The corresponding reference foils to aid in energy alignment and normalization were collected simultaneously with ex-situ measurements. For operando experiments, reference foil data were collected before and after operando measurements. HERFD-XAS data were collected at Cu K-edge, spectrometer was calibrated at Cu-K$\alpha$1 emission line (8047.8 eV) using Si (444) crystal. Additional details can be found in Supplementary Information.

**Electrochemical $CO_2RR$ measurements in H-cells**

Electrochemical $CO_2RR$ testing in H-type cells was performed by a custom-made three-electrode H-type cell system. Pt plate and Ag/AgCl (3 M KCl) were used as the counter and the reference electrodes, respectively. Cathode and anode chambers were separated with anion-exchange membrane (Selemion AMV-N, Asahi Glass) and filled with 1.8 mL of $CO_2$-saturated 0.1 M $KHCO_3$ (99.95%, Sigma Aldrich) solution. The $CO_2RR$ was carried out by applying step-wised potential from -1.5 V to -2.5 V (vs. Ag/AgCl) with a 0.5 V interval for 30 minutes. After $CO_2RR$ at each potential, 0.2 mL of catholyte and anolyte was collected and mixed with 0.1 mL of $D_2O$ (99.9 atom% D, Sigma Aldrich), containing 10 mM of dimethyl sulfoxide (DMSO) as an internal standard for quantification of liquid products by the proton nuclear magnetic resonance ([1]H-NMR,

Bruker Ascend 500 MHz instrument). Additional details can be found in Supplementary Information.

## Electrochemical CO₂RR measurements in MEA

A single MEA electrolyzer hardware (Fuel Cell Technology, FCT) with a single serpentine channel graphite flow field on the cathode and a single serpentine channel Pt-coated Ti flow field on the anode was used for $CO_2RR$ measurement. The cell was assembled by stacking as-prepared anode, membrane, and cathode and subsequently compressed to 40 in-lbs in 10 in-lbs increments to form an MEA configuration. 20% and 0% compression of the cathode GDL and anode PTL was achieved by controlling the ethylene tetrafluoroethylene (ETFE) gasket thickness. During the CO₂RR test, humidified CO₂ gas (99.999%, Airgas) at room temperature was continuously passed the cathode flow field at 200 cc min⁻¹, and 0.1 M KHCO₃ electrolyte was circulated with a rate of 20 mL min⁻¹ using peristaltic pump. The gas outline was connected to gas chromatography (SRI Company) through a water trap containing 20 mL of deionized water. Anolyte volume was confined to 20 mL like a water trap. The CO₂RR was carried out by applying step-wised cell potential from 3 V to 4 V in 0.5 V increments. Gas chromatography data were collected after 5 and 20 minutes of applying each potential. After 30 minutes of each potential application, 0.2 mL of sample in water trap and anolyte was collected and liquid products were quantified by ¹H NMR similar to the H-type cell experiment. FE of each product was calculated using the equation noted above. Additional details can be found in Supplementary Information.

## Computational Methods

Spin-polarized DFT calculations were performed using the Vienna Ab-initio Simulation Package (VASP) with a plane-wave basis set. Perdew-Burke-Ernzerhof (PBE) generalized gradient approximation (GGA) was used to describe the electron–electron exchange and correlation interactions. The DFT+U method was employed with U = 7 eV and J = 0 eV. The optimized the structural parameters using the bulk CuO model are a = 0.449 nm, b = 0.367 nm, c = 0.512 nm, $\beta$ = 96.1°, respectively. Additional details can be found in Supplementary Information.

## Data availability

All relevant data are available from the corresponding author on request.

## Acknowledgements

This work was supported by the U.S. Department of Energy, Office of Science, Office of Basic Energy Sciences, Materials Sciences and Engineering Division under Contract No. DE-AC02-05-CH11231 within the in-situ TEM program (KC22ZH). This work was supported in part by previous breakthroughs obtained through the Laboratory Directed Research and Development Program of Lawrence Berkeley National Laboratory (LBNL) under U.S. Department of Energy Contract No. DE-AC02-05CH11231. Portions of this work were supported by the Liquid Sunlight Alliance, which is supported by the U.S. Department of Energy, Office of Science, Office of Basic Energy Sciences, Fuels from Sunlight Hub under Award No. DE-SC0021266 as well as a CRADA with Chevron Corporation under Contract No. AWD00005793. This research used resources of the National Synchrotron Light Source II, a U.S. Department of Energy, Office of Science User Facility operated by Brookhaven National Laboratory under Contract No. DE-SC0012704. This work used the computational resources from the Extreme Science and Engineering Discovery Environment (XSEDE) through allocation DMR110087, which is supported by Department of Defense grant number N00014-19-1-2376. Work at the Molecular Foundry (LBNL), Cal-Cryo facility (Berkeley QB3 Institute), and Berkeley Marvell NanoLab was supported by the Office of Science, Office of Basic Energy Sciences, of the U.S. Department of Energy under Contract No. DE-AC02-05CH11231.

## Author contributions

J.W., E.L. and W.C. designed the experiments under the guidance of H.Z., W.S.D., and A.T.B.; J.W. and Y.C. synthesized CuO nanocatalysts, with the help of X.S., J.Y. and J.S.; J.W. performed electrochemical liquid cell fabrication, with the help of Y.C; J.W. performed quasi-operando EC-TEM measurements and performed analysis, with the help of X.S., K.C.B. and P.E.; J.W. performed ex situ TEM characterizations, with the help of Y.C. and Q.Z.; E.L. and J.W. performed operando XAS measurements and E.L. performed analysis under the guidance of D.L. and W.S.D.; W.C. and K.K. performed $CO_2RR$ performance measurements (H-type cell and MEA), with the help of E.L., W.L. and J.W. under the guidance of A.Z.W. and A.T.B.; H.X., M.Z. performed electron tomography measurements and analysis, with the help of J.W.; B.Z. performed DFT calculations under the guidance of M.A.; J.L performed SEM and AFM measurements under the guidance of Z. Y. A. B.; J.L. and J.W. prepared the scheme. J.W., E.L., W.C., J.L, B.Z. and Y.C. wrote the manuscript under the supervision of W.S.D. and H.Z. All authors revised and approved the manuscript.

**Competing interests**

The authors declare no competing interests.